# Z-source Inverter Based Grid-interface For Variable-speed Permanent Magnet Wind Turbine Generators

D. Mahinda Vilathgamuwa, *Senior Member, IEEE*; Xiaoyu Wang, *Member, IEEE*; King Jet Tseng, *Senior Member, IEEE*; C. J. Gajanayake, *Member, IEEE*

***Abstract*-- A Z-source inverter based grid-interface for a variable-speed wind turbine connected to a permanent magnet synchronous generator is proposed. A control system is designed to harvest maximum wind energy under varied wind conditions with the use of a permanent magnet synchronous generator, a diode-rectifier and a Z-source inverter. Control systems for speed regulation of the generator and for DC- and AC- sides of the Z-source inverter are implemented. Laboratory experiments are used to verify the efficacy of the proposed approach.**

## I. Introduction

RECENT trends in meeting increasing energy demand are moving towards generating power with distributed energy resources, and most of them are renewable as they have greater advantages due to their environmentally friendly nature and their ability of on-site generation. Furthermore, reliability of service and power quality are enhanced by proximity to the customer. This concept is commonly known as distributed generation (DG) and it has gained acceptance over the years. The DG complements centralized generation by having a relatively inexpensive response to incremental increases in power demand, by preventing existing transmission and distribution capacity upgrades, by locating power where it is most desired and by having the flexibility to put power back into the grid at user sites. Moreover, there are social demands for cheaper, less polluting, safer and more reliable and sustainable energy for consumers, suppliers, generators and policy makers. The DG, including integration of renewable sources, is a promising solution to solve those demands.

However, such renewable sources are distributed and technologies and standards necessary for their grid interconnection and isolated operation are presently being developed. There are numerous types of DG energy sources producing electrical power at different voltages and frequencies. Such disparity in output characteristics can be found even in sources like wind turbines, fuel cells and solar cells, all of them have a high potential to be dominant DG sources in future. Hence, it is necessary to convert the output voltage and frequency of such sources to standard values to be compatible with domestic and industrial loads. This can be achieved with a power conditioner.

Traditionally, there are two types of inverters used; commonly known as voltage source inverter (VSI) and current source inverter (CSI). Both of these inverters have limited operating range even though both are used in DG applications [1, 2]. To overcome the limited operating range, both these inverters need to be connected with a separate DC-DC converter stage at the front end. This enables them to operate in both buck and boost modes. This topology is commonly known as a two-stage inverter. Small scale two-stage inverters have been developed for domestic DG applications with fuel cells [3, 4]. However, two-stage inverters are not cost effective and also controlling them is known to be cumbersome. As a solution to this problem, Z-source inverter was proposed recently [5]. This is a single-stage inverter and it can operate in both buck and boost modes. The buck-boost characteristic is achieved due to the unique impedance network interfacing the inverter with the DC source. Moreover, it has better EMI properties due to the absence of the dead-time.

Amongst sustainable resources, wind energy has gained rapid development and has made a significant in-road into electrical power systems as a potential source of bulk power generation. Wind energy is derived from small area harnessing schemes as well as large wind farms that could generate significant amount of energy to loads [6-8]. Unfortunately, much like other renewable sources such as solar, wind generation tends to be unsteady because wind speed is influenced by natural and meteorological conditions. Notwithstanding such drawbacks, wind energy has been found to be a technically and economically feasible option for generating power. With the advent of high speed and efficient power electronic devices and variable-speed direct-driven generators, a quiet and economical wind generation system has become a reality.

Synchronous generators have been used for direct-coupled and low-speed wind generation applications. Particularly,

D. Mahinda Vilathgamuwa, King Jet Tseng, and C. J. Gajanayake are with School of Electrical and Electronic Engineering, Nanyang Technological University, Block S2, Nanyang Avenue, Singapore 639798 (e-mail: emahinda@ntu.edu.sg, ekjtseng@ntu.edu.sg, and jgchandana@ntu.edu.sg).

X. Y. Wang is with Sustainable Energy Technologies Department of Brookhaven National Laboratory, Upton, NY 11961, USA. (e-mail: xywang@bnl.gov ).

permanent magnet (PM) type synchronous generators have been gaining acceptance for such applications recently as they are highly efficient and are of relatively smaller in diameter. The AC voltage produced by PM generators can be rectified to generate a smooth DC voltage using a simple diode rectifier to reduce the cost. The output of the combination of PM generator and diode rectifier is uncontrollable in nature as PM generators lack excitation control. Moreover, the shaft speed of the generator needs to be varied in proportion to the wind speed to keep the "tip-speed ratio" of the turbine at its optimum value that corresponds to the maximum power generation [7]. As the diode rectifier is uncontrollable, the grid interfacing inverter current is controlled to draw a proportional current from Permanent Magnet Synchronous Generator (PMSG) so that a torque necessary for tracking the optimal speed of PMSG is produced eventually. With the variation of shaft speed of PMSG, the emf produced by PMSG and consequently the rectified DC voltage would vary with time. To interface such widely varying DC voltage to the mains grid, a Z-source inverter based single converter stage is proposed in this paper. The proposed topology is economical and less complex compared to traditional two-stage converter topologies. The control methodologies of the proposed converter topology are presented along with simulation and experimental results to show the efficacy of the proposed approach.

## II. System Configuration

The proposed Z-source inverter based PMSG wind turbine system is illustrated in Fig. 1. In the figure, the PMSG converts the wind turbine captured wind power to electrical form, which is denoted by $P_w$. As wind speed is intermittent in nature, $P_w$ would vary with time [8]. The front-end converter rectifies the generated AC power output into a DC voltage, $v_{dc}$ across the DC-link capacitor, $C_1$. The fluctuation in wind power results in the variation of the DC-link voltage. To convert the variable DC voltage into an AC form with specified voltage and frequency, the Z-source inverter is proposed as shown in Fig. 1. The Z-source inverter consists of Z-source impedance network connected to DC-side of a standard PWM voltage source inverter. With a suitably designed controller, the Z-source inverter can operate under varied DC-link voltage conditions while maintaining the magnitude of inverter output voltage, $v_o$ constant.

## III. Controller Design

The control system of the Z-source inverter based wind power system is shown in Fig. 2. The overall system comprises of the DC-side control loop and the AC-side control loop, as shown in green- and red-boxes respectively. In the following sub-sections, the details of the two controllers are discussed in detail.

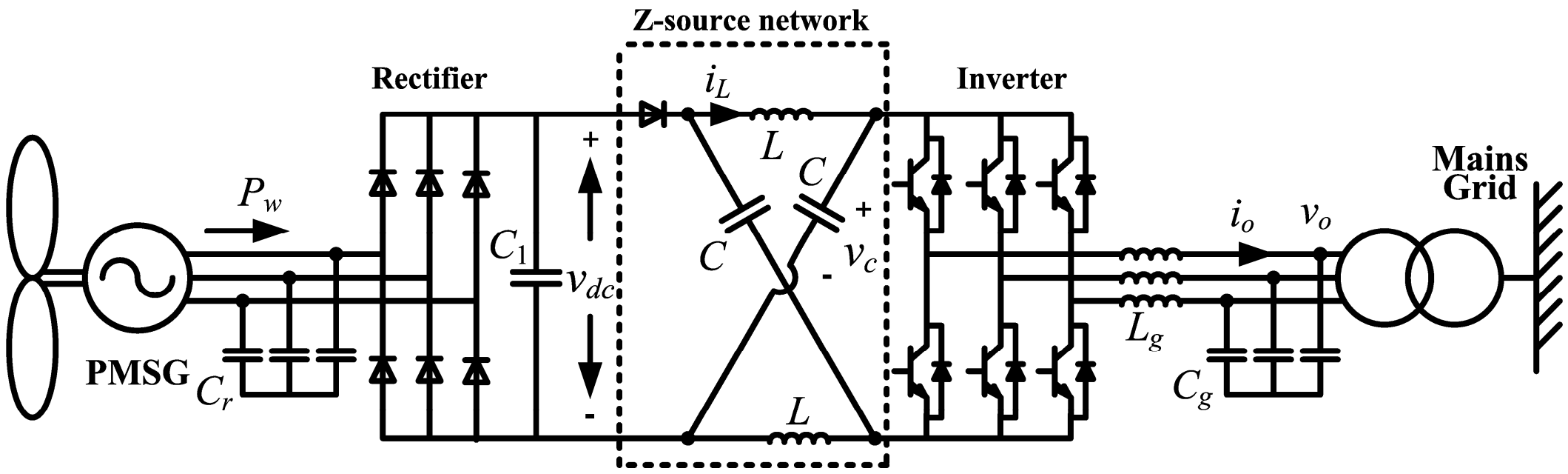

Fig. 1. Wind power generation scheme with Z-source inverter grid interface.

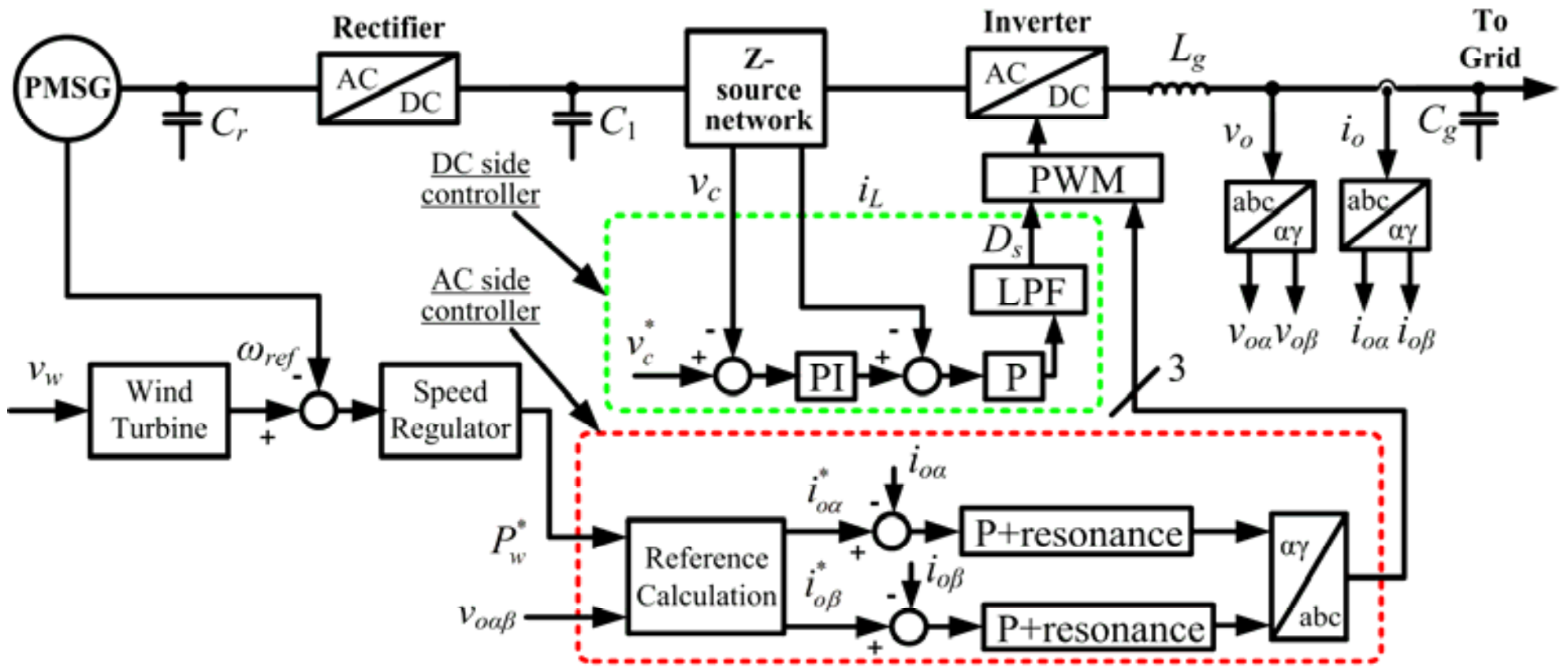

Fig. 2. Control scheme of the Z-source inverter based wind power generation system.

*A. Derivation of reference power component for optimal wind power generation*

The mechanical power $P_w$ transmitted to the wind turbine shaft is given by (1) where $\rho$ is density of air, $A$ is area swept by blades, $V_w$ is wind speed and $C_p$ is turbine power coefficient. When the pitch angle of the blade is constant, the turbine power coefficient $C_p$ is a function only of the tip speed ratio λ, which is the ratio of the wind turbine blade tip speed to the wind speed $V_w$ as given in (2). In this equation, $R$ is the radius of the blades and $\omega_w$ is the angular speed of the wind turbine shaft.

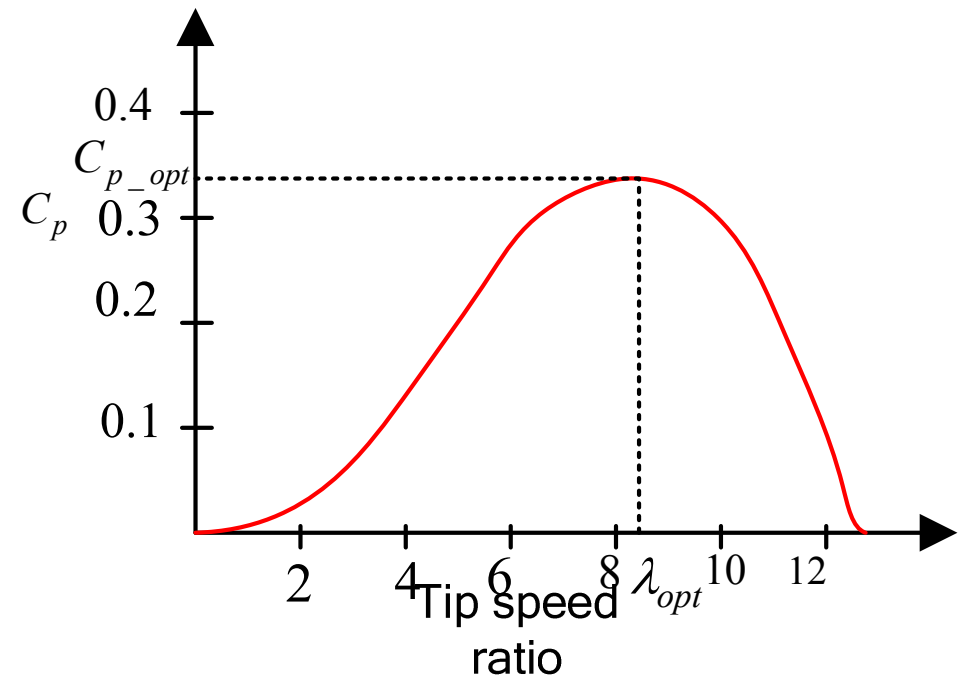

Fig. 3. Variation of power coefficient $C_p$ with tip speed ratio.

The variation of the power coefficient with the tip speed ratio for a typical wind turbine is shown in Fig. 3. When $\lambda$ takes the optimal value $\lambda_{opt}$, the power coefficient $C_p$ becomes maximum at $C_{p,opt}$. When the tip speed ratio λ is maintained at its optimum value $\lambda_{opt}$ regardless of the wind speed, the maximum mechanical power can be derived from wind. The maximum power that is derived from the wind turbine $P_{w,opt}$ is given in (3) and $P_{w,opt}$ is proportional to the cubic power of the turbine shaft speed. In this optimum condition, the generator speed is kept proportional to the wind speed, as shown in (4) where $K_\omega = \lambda_{opt} /R$. Therefore, the desired generator speed $\omega_{ref}$ is sent to a speed regulator to obtain a value for the reference power $P_w^*$ which in turn is used to derive reference current components for the grid inverter current control. It can be seen that if the speed regulator maintains the generator speed at $\omega_{ref}$, $P_{w,opt}$ is transferred to the grid through power flow control of the grid-side inverter.

$$P_w = 0.5\rho A V_w^3 C_p \tag{1}$$

$$\lambda = \frac{R\omega_w}{V_w} \tag{2}$$

$$P_{w,opt} = \frac{0.5\rho A \omega_w^3 R^3 C_{p,opt}}{\lambda_{opt}^3} \tag{3}$$

$$\omega_{ref} = K_\omega V_w \tag{4}$$

*B. DC-side controller design*

Similar to the controller design proposed in [9], the Z-source inverter in the proposed scheme is modeled as two cascaded sub systems with time scale decoupling between them. The controllers are designed independently. The AC-side voltage disturbances would have minimum effect on the DC-side. However, the changes in grid current could alter the inductor current of the Z-source impedance network. Grid current variations can be considered as a disturbance in the AC-side and can be compensated for. However, to prevent the clashes between the dynamics of AC- and DC-sides, the DC-side dynamics should be made considerably slower. This could be supported by having a higher bandwidth in the AC-side voltage and current loops. Small signal analysis is performed to obtain a linear model of the Z-source impedance network. Its block diagram representation is shown in Fig. 4, which shows the relationship between state variables and shoot-through duty ratio, where $V_{11}=2V_C-V_{DC}-RI_{DC}$ and $I_{11}=2I_L-I_{DC}$, $D_S$ = shoot-through state duty-ratio, $D_A$= non-shoot-through state duty-ratio [9]. In contrast to the controller design proposed in [9], controllers here are designed to control the voltage across the Z-source capacitor, due to two reasons. First, the output voltage of Z-source impedance network is pulsating and secondly, this would allow the switches to have minimum voltage stress under varying conditions and the boosted voltage could be effectively utilized.

From Fig. 4, the transfer function of $\tilde{V}_C/\tilde{D}_S$ is derived as shown in (5), however, it has a RHP zero. This is a clear indication of the presence of non-minimum phase. Hence, the design of closed-loop controllers should be carried out

$$\frac{\tilde{V}_C}{\tilde{D}_S} = \frac{2V_C - V_{DC} - RI_{DC} - (2I_L - I_{DC})(R+r) - L(2I_L - I_{DC})s}{LCs^2 + (R+r)Cs + (D_A - D_S)^2} \tag{5}$$

$$\frac{\tilde{I}_L}{\tilde{D}_S} = \frac{\left(2V_C - V_{DC} - RI_{DC}\right)Cs}{\left(LCs^2 + (R+r)Cs + (D_A - D_S)^2\right)} \tag{6}$$

$$\frac{\tilde{V}_C}{\tilde{I}_L^*} = \frac{K_{Lp}\left((2V_C - V_{DC} - RI_{DC})(D_A - D_S) - (2I_L - I_{DC})(R+r) - L(2I_L - I_{DC})s\right)}{LCs^2 + \left(R + r + 2K_{Lp}\left(V_C - V_{DC} - RI_{DC}\right)(D_A - D_S)\right)Cs + (D_A - D_S)^2} \tag{7}$$

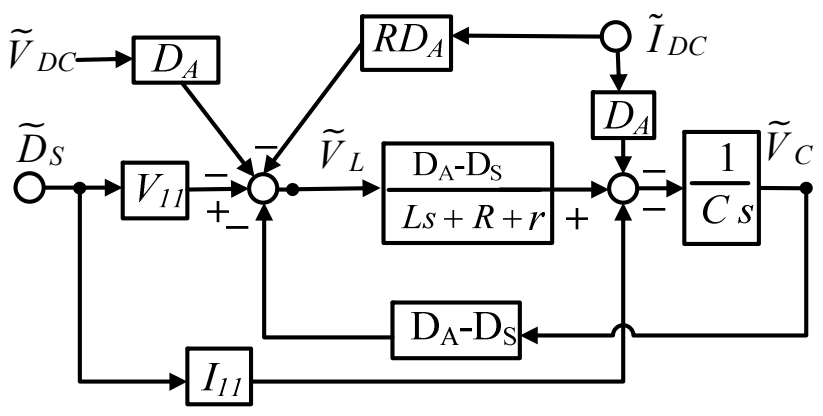

Fig. 4. Block diagram representation of Z-source impedance network.

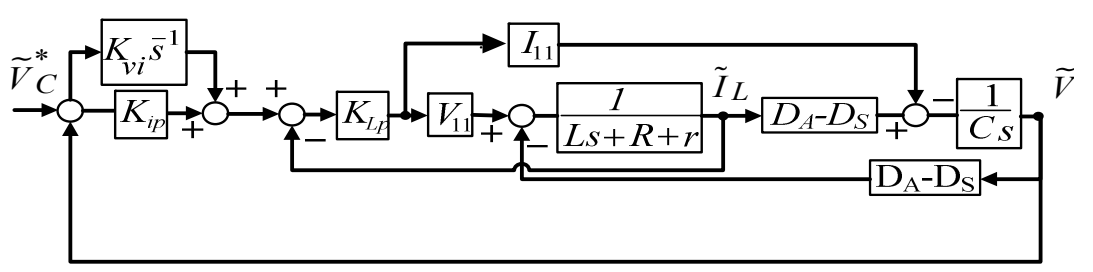

Fig. 5. Block diagram representation of Z-source impedance network with closed current and voltage loops.

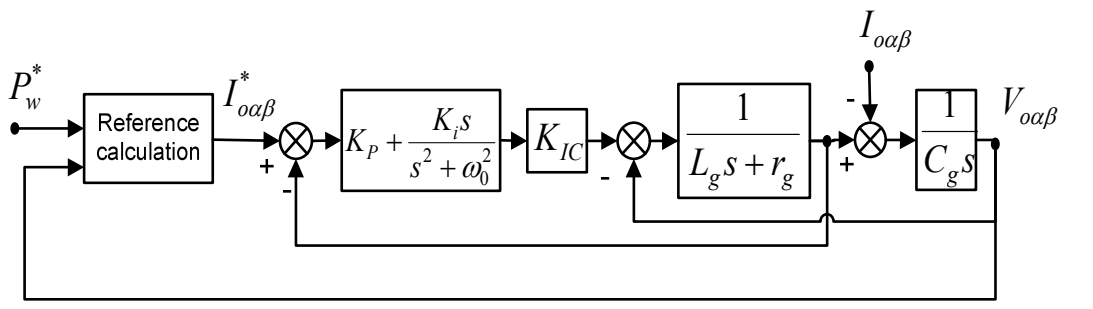

Fig. 6. Block diagram of the AC side closed loop controller.

carefully. The non-minimum phase problem is handled with having two loops, i.e. inner current and outer voltage loops, and the technique is commonly known as indirect controller. A similar technique is proposed here to obtain a stable feedback controller. From Fig. 4, the open-loop transfer function of the inner current loop can be derived and is given in (6). It has no RHP zero making inner loop design less tedious and therefore a proportional controller is employed as shown in Fig. 5. Subsequently, the open loop transfer function for the outer capacitor voltage loop is derived as in (7) and then to achieve the required bandwidth and to maintain stability, a PI controller is cascaded. Selected parameters prevent clashes between the dynamics of AC- and DC-sides, as the crossover frequency of the DC-side outer-loop is made very much smaller than that of the outer voltage loop in the AC-side.

### C. AC-side controller design

Consider the AC-side of Fig. 1 and by applying KCL and KVL, the mathematical model of the AC-side can be derived. The block diagram representation of the AC-side is given in Fig. 6. The AC-side output current is controlled by changing the modulation index. The output inverter inductor current measurement is used as the feed-back signal. From Mason's gain rule, it is possible to obtain the open-loop transfer functions for $\alpha\beta$ axes and controllers are designed using the obtained transfer functions. Recently proposed P+resonance controller is employed to track the current reference [10]. This controller performs like a PI controller designed in rotating reference frame but without the computational burden of transformation into the rotating reference frame.

In this paper, the aim of the controller design is to deliver power generated by the PMSG. This is achieved by injecting the required current into the grid. However, with the changes in grid voltage and in wind speed, the current reference needs to be changed continuously. The reference currents are generated using instantaneous power theory [11] and are given in (8) and (9) in the $\alpha\beta$ reference frame. $P_w^*$ is the power reference from the PMSG regulator and $v_{o\alpha}$ and $v_{o\beta}$ are the measured grid voltage components in the stationary reference frame.

$$I_{o\alpha}^* = \frac{V_{o\alpha}}{V_{o\alpha}^2 + V_{o\beta}^2} P_w^* \tag{8}$$

$$I_{o\beta}^* = \frac{V_{o\beta}}{V_{o\alpha}^2 + V_{o\beta}^2} P_w^* \tag{9}$$

## IV. EXPERIMENTAL VERIFICATIONS

To verify the effectiveness of the proposed control schemes, the prototype Z-source inverter based wind integration system is built in the laboratory. The developed prototype is a scaled down model of the actual system and is shown in Fig. 1 with the exception that a DC machine is used to emulate the wind turbine as the prime mover of PMSG. A three phase AC power supply connected to a resistive load bank is used to emulate the grid. The experimental system parameters are given in Table I. Figs. 7 and 8 show the experimental results obtained from the prototype.

Table I: Parameters for the experiments

| Parameter | Value |
|---|---|
| L | 5 mH |
| C | 2200 μF |
| $C_1$ | 6800 μF |
| $L_g$ | 5 mH |
| $C_g$ | 25μF |
| $R_{load}$ | 25Ω |
| M- max | 0.9 |
| $D_s$-max | 0.35 |
| DC machine, $P_{rated}$ | 1.2 kW |
| DC machine, $\omega_{rated}$ | 1400 rpm |
| DC machine stator, $V_{rated}$ | 220 V |
| DC machine stator, $I_{rated}$ | 6.0 A |
| DC machine exciter, $V_{rated}$ | 220 V |
| DC machine exciter, $I_{rated}$ | 0.55 A |
| PMSG, $P_{rated}$ | 1.5 kW |
| PMSG, $V_{rated}$ | 400 V |
| PMSG, Poles | 4 |
| PMSG, $\omega_{rated}$ | 1500 rpm (@ 50 Hz) |

The experimental verifications of the AC- and DC-side controllers of the Z-source inverter have been done and the results are given in Figs. 7 and 8 respectively. In the experiment, the Z-source impedance network capacitor reference voltage is kept constant and step changes in the PMSG reference speed are carried out. The waveforms of the Z-source impedance network capacitor voltage, input DC voltage, output voltage and output current during the speed step-down and step-up transitions are shown in Figs. 7(a) and (b) respectively. From Fig. 7(a), one can find that, with the speed reference is reduced from 53 rad/s to 50 rad/s, the output power from the DC machine is increased as prime mover approaches its maximum power with the decrease of the speed and accordingly, the AC-side output current is stepped-up. Though the input DC voltage decreases due to the PMSG speed is reduced, the Z-source impedance network capacitor voltage is well controlled and kept at its original value over this transition interval. The waveforms reveal that, with the reference speed changes back from 50 rad/s to 53 rad/s, the AC-side output current is correspondingly restored to its original value. The capacitor voltage is again well controlled when the input DC voltage is reduced with the reduction of the PMSG speed. The AC output voltage- and current-waveforms in the high speed (53 rad/s) and low speed (50 rad/s) conditions are highlighted in Figs. 7(c) and (d) respectively.

The second experiment is carried out to verify the effectiveness of the DC side controller. In this experiment, the PMSG reference speed is kept unchanged at 53 rad/s, while the Z-source impedance network capacitor voltage reference is changed from 140 V to 170 V and back to 140 V. The waveforms of the Z-source impedance network capacitor voltage, input DC voltage, output voltage and output current during the step transitions are shown in Figs. 8(a) and (b). At the original condition, the input DC voltage is sufficient to produce the desired Z-source impedance network capacitor voltage without shoot-through. With the capacitor voltage reference is increased to 170 V, shoot-through intervals need to be inserted to boost up the Z-source impedance network capacitor voltage while the DC input voltage is maintained constant with the PMSG speed is controlled unchanged. In this situation $D_{sh}$ for the Z-source is found to be 0.2 with modulation index = 0.3. In the original condition, modulation index = 0.6 with $D_{sh}$ = 0. When the capacitor voltage reference is reduced to 140 V, the capacitor voltage decreases as shown in Fig. 8(b). The DC input voltage remains unchanged, revealing the fact that the PMSG speed is well controlled. The waveforms of the AC-side voltage and current for capacitor voltage reference settings of 140 V and 170 V are given in Figs. 8(c) and (d) respectively.

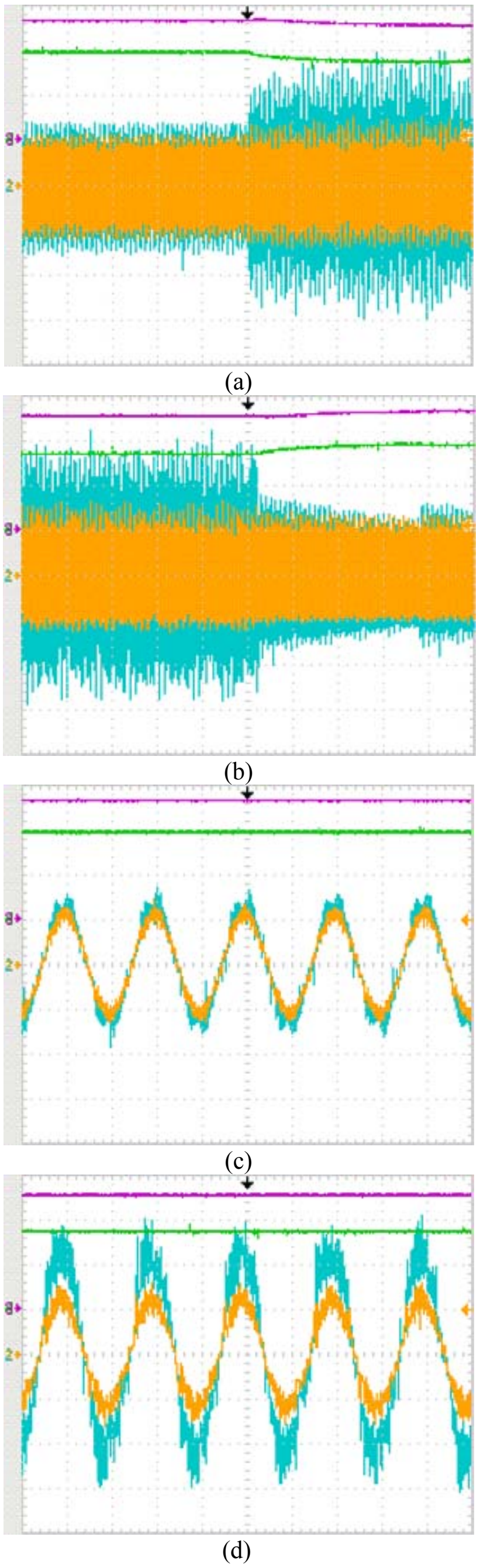

Fig. 7. Experimental results, (a) for a reduction in reference speed, waveforms of the Z-source impedance network capacitor voltage 50 V/div, input DC voltage 50 V/div, output voltage 10 V/div and output current 1 A/div (from top to bottom), (b) for an increase in reference speed, waveforms of the Z-source impedance network capacitor voltage 50 V/div, input DC voltage 50 V/div, output voltage 10 V/div and output current 1 A/div (from top to bottom), (c) AC output voltage 10 V/div (orange) and current 1 A/div (blue) waveforms at the speed of 53 rad/s, and (d) AC output voltage 10 V/div (orange) and current 1 A/div (blue) waveforms at the speed of 50 rad/s.

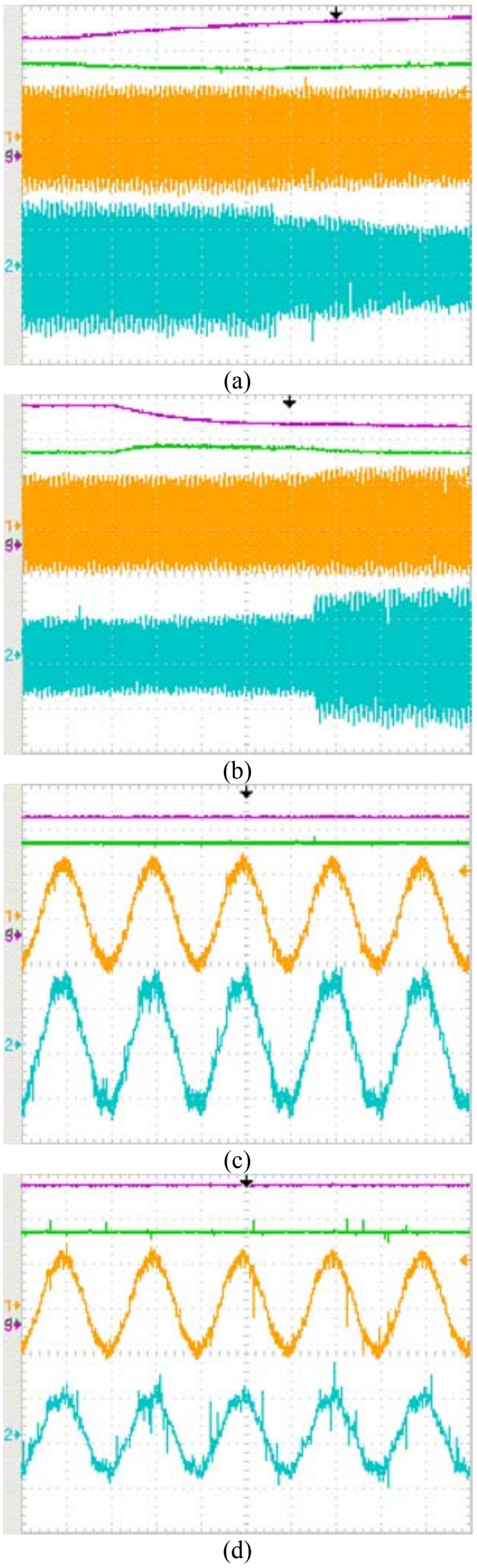

Fig. 8. Experimental results, (a) for a Z-source capacitor voltage reference increase, waveforms of the Z-source impedance network capacitor voltage 50 V/div, input DC voltage 50 V/div, output voltage 10 V/div, and output current 1 A/div (from top to bottom), (b) for a Z-source capacitor voltage reference decrease, waveforms of the Z-source impedance network capacitor voltage 50 V/div, input DC voltage 50 V/div, output voltage 10 V/div, and output current 1 A/div (from top to bottom), (c) AC output voltage 10 V/div (orange) and current 1 A/div (blue) waveforms at a capacitor voltage of 140 V, and (d) AC output voltage 10 V/div (orange) and current 1 A/div (blue) waveforms at a capacitor voltage of 170 V.

## V. Conclusions

A wind power integration topology based on Z-source inverter system is proposed in this paper. With the employment of the proposed Z-source inverter, the AC-side voltage is maintained constant though the DC-link voltage of the front end rectifier tends to fluctuate due to the stochastic nature of the wind. The DC- and AC-side controllers for the Z-source inverter are designed. The DC-side controller consists of two cascaded sub-systems which are designed with the use of time scale decoupling between them. With the use of suitably selected PI constants, the DC-side voltage disturbances would have minimum effect on the AC-side. The recently proposed P+resonance controller is applied to the AC control system. The output current is controlled to track the fluctuating wind power generated from the PMSG. The effectiveness of the proposed controllers has been verified by experiment results.